\documentclass{iaus}
\usepackage{graphicx}

\title[IAUS 265.~~Stellar Parameters of Stars with Planets] 
{Stellar Parameters for a Sample of Stars with Planets}

\author[Ghezzi et al.]   
{Luan Ghezzi$^1$, Katia Cunha$^{2,1}$, Francisco X. de Ara\'ujo$^1$, Verne V. Smith$^2$, Ramiro de la Reza$^1$, 
and Simon Schuler$^2$}

\affiliation{$^1$Observat\'orio Nacional, Rua General Jos\'e Cristino, 77, 20921-400, \\
                 S\~ao Crist\'ov\~ao, Rio de Janeiro, RJ, Brazil; email: {\tt luan@on.br} \\[\affilskip]
             $^2$NOAO, 950 N Cherry Ave., Tucson, AZ 85719, USA \\[\affilskip]}

\pubyear{2009}
\volume{265}  
\pagerange{119--126}
\jname{Chemical Abundances in the Universe: Connecting First Stars to Planets}
\editors{K. Cunha, M. Spite \& B. Barbuy, eds.}
\begin{document}

\maketitle

\begin{abstract}

The study of chemical abundances in stars with planets is an important
ingredient for the models of formation and evolution of planetary systems.
In order to determine accurate abundances, it is crucial to have a reliable 
set of atmospheric parameters. In this work, we describe the 
homogeneous determination of effective temperatures, surface gravities 
and iron abundances for a large sample of stars with planets as well
as a control sample of stars without giant planets. Our results indicate 
that the metallicity distribution of the stars with planets is more metal 
rich by $\sim$ 0.13 dex than the control sample stars.




\end{abstract}

\firstsection 
\section{Introduction}

More than 400 stars with planets have been detected up to this date. 
One of the interesting properties of these stars concerns their metallicity distribution.
Several studies (e.g., \cite[Fischer \& Valenti 2005]{fv05}) have
confirmed the result first shown by \cite{g97}: stars with giant planets are
systematically metal-rich (by $\sim$ 0.2 dex) relative to field FGK dwarfs not known to
harbor planets. Two hypotheses have been proposed to explain this excess:
primordial enrichment or pollution. Current results (e.g., \cite[Fischer \& Valenti 2005]{fv05}) 
show that the frequency of planets 
increases significantly for higher metallicities, thus giving strong support for the primordial 
hypotheses. Evidence for the occurence of pollution is still ambiguous (for a comprehensive review, 
see \cite[Gonzalez 2006]{g06}). 


\section{Observations and Data Reduction}


We have observed a sample of 156 and 160 stars with and without planets, respectively. Spectra 
of the program stars were obtained at the MPG/ESO-2.20m telescope (La Silla, Chile) with the FEROS 
spectrograph (under the agreement ESO-ON). They have an almost complete spectral coverage from 
3500 to 9200 \AA, R $\sim$ 48.000 and S/R $>$ 200 $–-$ 300 (at $\sim$ 6300 \AA). The data were 
reduced with the FEROS-DRS package. It is worth noting that 6 stars of this sample have already been 
analyzed for $^{6}$Li in \cite{ghezzi09}, showing no evidence of pollution.

\section{Determination of the Stellar Parameters}

Stellar parameters for all the target stars were derived spectroscopically and followed standard techniques. 
Effective temperatures were obtained from zero slopes in diagrams of A(Fe I) \textit{versus} $\chi$ 
using just lines which had $\log (EW/\lambda) < -5.00$. This limit was set to eliminate the strong lines 
that are sensitive to the microturbulent velocity value, decoupling the $T_{eff}$ determination from the microturbulence determination. The microturbulent velocities were varied until the slopes of A(Fe I) \textit{versus} 
$\log (EW/\lambda)$ were zero. Finally, surface gravities were derived from ionization equilibrium 
between Fe I and Fe II species. The abundances were derived in LTE using an updated
version of the spectrum synthesis code MOOG (\cite[Sneden 1973]{s73}). The model atmospheres
adopted in the analysis were interpolated from the ODFNEW grid of ATLAS9 models 
(\cite[Castelli \& Kurucz 2004]{ck04}).

\section{Results and Next Steps}

\begin{figure}
\begin{center}
 \includegraphics[width=6cm]{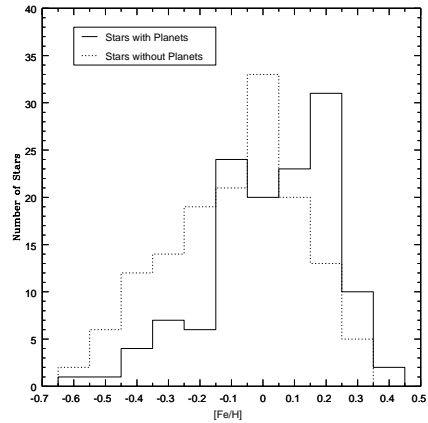} 
 \caption{The metallicity distributions for a sample of 125 planet host stars (solid line) and a 
          control sample with 149 stars without giant planets.}
   \label{fig1}
\end{center}
\end{figure}

The average metallicities for stars with and 
without planets are +0.02 and -0.09, respectively (Fig.\,\ref{fig1}). These results recover the offset 
between the average
metallicities of the two samples ($\sim$ 0.13 dex) as well as the average metallicity for field FGK 
dwarfs usually found 
in the literature. The next step of this study is the homogeneous determination of the abundances of 
other chemical elements 
(such as Ni, Si, Ti, V, Ca, Na, C, N, O, Mg and Li) for the entire sample. The comparison 
of the abundance patterns for stars with and without planets (for instance, using the relation between 
chemical abundances of several elements and their condensation temperatures; see e.g., 
\cite[Smith, Cunha \& Lazzaro 2001]{s01} and \cite[Ecuvillon et al. 2006]{e06}) will allow us to check 
the occurrence of the pollution process and understand better the formation and evolution of 
planetary systems.

\end{document}